\documentclass[aps,twocolumn,a4paper,pra,superscriptaddress,floatfix,showpacs]{revtex4}%
\usepackage{amssymb}
\usepackage{amsmath}
\usepackage{amsfonts}
\usepackage{graphicx}
\usepackage{graphics}

\usepackage{dcolumn}
\usepackage{bm}

\begin{document}

\title{Driving quantized vortices with quantum vacuum fluctuations}

\author{Fran\c cois Impens}
\author{Ana M. Contreras-Reyes}
\author{Paulo A. Maia Neto}
\affiliation{Instituto de Fisica, UFRJ, CP 68528, Rio de Janeiro, RJ, 21941-972, Brazil}
\author{Diego A.R. Dalvit}
\affiliation{Theoretical Division, MS B213, Los Alamos National Laboratory, Los Alamos, NM 87545, USA}
\author{Romain Gu\'erout}
 \author{Astrid Lambrecht}
 \author{Serge Reynaud}
\affiliation{Laboratoire Kastler Brossel, case 74, CNRS, ENS, UPMC, Campus Jussieu, F-75252 Paris Cedex 05, France}



\begin{abstract}
We propose to use a rotating corrugated material plate in order to stir, through the Casimir-Polder interaction, quantized vortices in an harmonically trapped Bose-Einstein condensate. The emergence of such vortices within the condensate cannot be explained with a computation of the Casimir-Polder potential based on the pairwise summation approach or on the proximity force approximation. It thus appears as a genuine signature of non-trivial geometry effects on the electromagnetic vacuum fluctuations, which fully exploits the superfluid nature of the sample. In order to discuss quantitatively the generation of Casimir-driven vortices, we derive
 an exact non-perturbative theory of the Casimir-Polder potential felt by the atoms in front of the grating. 
 Our numerical results for a Rb condensate close to a Si grating show 
 that the resulting quantum vacuum torque is strong enough to provide a contactless transfer of angular momentum to the condensate and generate quantized vortices under realistic experimental conditions at separation distances around $3\,\mu{\rm m}.$
 \end{abstract}

\maketitle

Since the advent of Bose-Einstein condensation~\cite{BECreview}, the specific features of ultra-cold atomic samples have turned them into a versatile tool useful well-beyond the field of laser cooling and in particular for precision experiments. The extremely narrow momentum distribution of such samples can indeed be used in atom interferometric systems serving as inertial sensors or atomic clocks~\cite{matterwaveresonators}. Recently, Bose-Einstein condensates (BECs) have attracted considerable attention as promising probes of the atom-surface Casimir-Polder interaction~\cite{CasimirPolder,CasimirReview09}. Casimir forces can indeed give rise to noticeable effects in the condensate dynamics, affecting for instance the stability of a magnetically trapped sample~\cite{Vuletic04}, or provoking quantum reflections when the BEC approaches a surface sufficiently slowly~\cite{QuantReflectionBEC}. To date, the more accurate measurements of the Casimir-Polder force
 performed with a BEC~\cite{CornellExperiments} rely on the tiny correction to the center-of-mass oscillation frequency induced by a nearby material plate~\cite{Antezza04}.

The experiments realized or proposed so far involve only the translational motion of the ultra-cold atom sample. We propose instead to probe Casimir forces through the contactless transfer of angular momentum from a rotating grating to a harmonically trapped condensate.
Quantum vacuum or Casimir torques have been predicted to arise between two gratings when their corrugations are not aligned~\cite{Rodrigues06} or between two flat but birefringent plates \cite{Munday05}. However they have not been measured experimentally so far. In this Letter, we take advantage of the cold atoms technology and propose a vacuum torque between a rotating grating and a BEC resulting in the excitation of quantized vortices inside the BEC.
This approach brings several benefits. It takes advantage of one of the most prominent features of superfluidity, namely the quantization of velocity circulation~\cite{RMPFetter}. In contrast to the experiments reported in Refs.\cite{CornellExperiments}, in our proposal the Casimir-Polder potential does not
induce a small correction onto the atomic motion, it rather plays the central part in the stirring process: in the absence of Casimir effect, the ultra-cold sample would acquire no angular momentum. The emergence of quantized vortices is thus a genuine signature of the atom-surface interaction mediated by the
quantum vacuum field.
A fortunate feature of BECs is their ability to be efficiently stirred by a slightly anisotropic rotating potential, enabling the production of one-vortex states through a weak time-dependent contribution~\cite{Madison00}. This property
 turns condensates into highly sensitive probes of rotating potentials. The purpose of this Letter is to show that, for realistic experimental parameters, the nucleation of quantized vortices can be driven by a rotating Casimir-Polder potential.

 In order to enhance the Casimir-induced anisotropy of the potential felt by the condensate, it is necessary to position it at a distance of the order of the corrugation period, which invalidates~\cite{Dalvit08,Dalvit08A,Messina09} the proximity force approximation (PFA)~\cite{PFA} and the pairwise summation approach (PWS)~\cite{PWS} usually employed to derive Casimir forces.
 Within PFA, the
  Casimir-Polder potential is approximated by the result for a plane interface, taken at the local atom-surface distance.
 As discussed below, it is precisely the departure from the PFA and PWS approximations that provides the dominant contribution to the time-dependent potential driving the vortices. In fact, neither the PFA nor the PWS treatments yield a sufficient torque to generate vortices in the sample.
  In this sense, the proposed setup probes non-trivial geometry effects through the nucleation of quantized vortices, which contrasts to previous experiments~\cite{CornellExperiments} measuring Casimir forces in a  simpler planar configuration.
  The Casimir-Polder potential for a grating could also be probed
  via Bragg spectroscopy of the energy spectrum
    \cite{Moreno10}  of a BEC, or by measuring its
   center-of-mass (dipole) lateral oscillation frequency~\cite{Dalvit08A}.
    We explore here how such geometry effects can excite modes of the atomic-field presenting a non-zero angular momentum.

  The Casimir-Polder
  potential for a corrugated surface has been computed beyond the PWS and PFA treatments up to first order in the corrugation amplitude~\cite{Dalvit08A,Messina09}. We develop here a more general non-perturbative method based on the scattering approach to dispersive interactions \cite{Lambrecht06,Emig07},
 suitable to address the larger amplitudes considered here (see also~\cite{Dobrich08} for a scalar field model). This formalism,
 similar to the theory developed in Ref.\cite{Lambrecht08} for treating surface-surface interactions, is built on the 
 analysis of the  scattering of vacuum field fluctuations  between the ground-state atom and the grating.  
 By taking the dispersive electromagnetic responses of both atom and material medium into account, it covers the entire range of distances, from the unretarded short-distance van der Waals regime to
 the Casimir-Polder large-distance asymptotic limit. For the relatively large distances considered here,  we are closer to the Casimir-Polder limit, but 
 our numerical results are nevertheless exact and take dispersion fully into account. 
  A detailed derivation of the dispersive potential is presented  elsewhere~\cite{AnaNonPerturbativeCasimir}.

\begin{figure}[t]
\centering
\includegraphics[width=8cm]{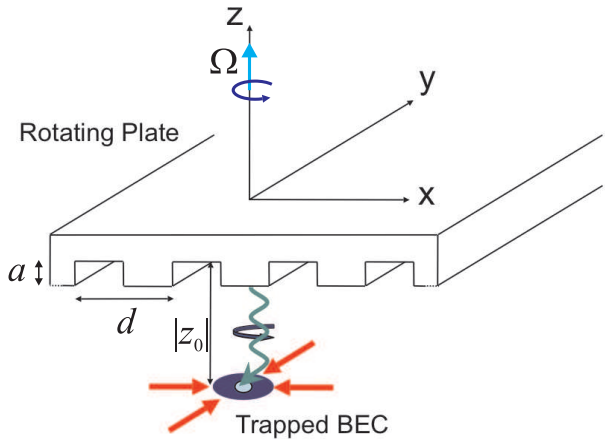}
\caption{Proposed experimental setup.}
\label{fig:rotating plate}
\end{figure}

 We consider the following setup, sketched in Fig.\ref{fig:rotating plate}.
  It consists in a time-independent harmonic optical trap with cylindrical symmetry around the vertical axis denoted as $O_z$, the center of which is situated close to and below a plate rotating at an angular velocity $\Omega$ around this axis. The trap center lies on the rotation axis. The plate is assumed to have a square corrugation profile of period $d$, and to be sufficiently large to be treated as infinite. The trap is loaded with a BEC preferably realized with an all-optical method~\cite{AllOpticalBEC} in order to avoid spurious magnetic effects such as the induction of local charges on the rotating plate. We assume the sample to be
well described within the mean-field approximation~\cite{BECreview}. We denote
$U_{0}(r,z)= \frac 1 2 m \omega_{0 \bot}^2 (x^2+y^2) + \frac 1 2 m \omega_{0 z}^2 (z-z_0)^2$ the sum of the gravitational and trapping potential and $U_{\rm CP}(r,z,t)$ the time-dependent Casimir-Polder potential induced by the plate,
where $z_0$ is the location of the trap center along the $z$-axis after taking into account the gravitational sag. The trap lateral position is at
$x=y=0.$
 The transformation of the Hamiltonian into the rotating frame can be written as $\widetilde{\hat{H}}=\hat{H}- \Omega \hat{L}_z$, where $\hat{H}$ and $\widetilde{\hat{H}}$ are the laboratory and rotating frame Hamiltonians respectively, while $\hat{L}_z$ is the vertical angular momentum. The second term in the r.h.s. favors the emergence of quantum states with an upward vertical angular momentum, which implies the presence of vortices in the considered superfluid sample.

We assume that the size of the condensate is sufficiently small to perform a quadratic approximation of the Casimir-Polder potential in the transverse coordinates with an effective frequency $\omega_{\rm CP}$ evaluated at the central sample vertical position $z_0:$
$U_{\rm CP}(x,z) \simeq U_{\rm CP}(0,z)+ \frac {1} {2} m \omega_{\rm CP}^2(z_0) x^2.$
 We have used the translational invariance of our setup along the axis $O_y$. These assumptions
  limit the transverse and longitudinal radius of the sample to respectively a quarter of the corrugation period $d$ and a fraction of the distance to the plate:
\begin{equation}
\label{eq:conditions sur les rayons}
R_{\bot} \leq \: c_{\bot} \: \frac {d} {4}, \quad \quad \quad R_{z} \leq c_{\!/\!/} \: z_0 \,,
\end{equation}
with $c_{\bot} \simeq 1$ and $c_{\!/\!/} \ll 1$ .
Considering a sample of small transverse size allows us to take advantage of the considerable theoretical and experimental work performed on the rotation of harmonically trapped condensates through time-dependent quadratic potentials~\cite{RMPFetter}. Besides, in this configuration the PFA treatment predicts a Casimir-Polder potential which is time-independent in the sample region and thus unable to transfer any angular momentum to the condensate (the PFA potential is piece-wise constant in the transverse coordinates with jumps following the plate discontinuities.).

The macroscopic sample wave-function then satisfies the time-dependent Gross-Pitaevskii equation (GPE) in the rotating frame~\cite{BECreview}
\begin{equation}
\label{eq:rotating frame equation}
i \hbar \frac {\partial \psi} {\partial t} = \left[ - \frac {\hbar^2} {2 m} \Delta+ U(\mathbf{r},t) + N g |\psi|^2 - \Omega \hat{L}_z \right] \psi \,,
\end{equation}
where the effective quadratic potential $U(\mathbf{r},t)$ becomes anisotropic under the influence of the Casimir-Polder interaction
\begin{eqnarray}
U(\mathbf{r},t) & = & \frac {1} {2} m \omega_{\bot}^2 \left[ \left( 1+ \epsilon \right) x^2+\left( 1- \epsilon \right)y^2+ \frac {\omega_{0z}^2} {\omega_{\bot}^2}
(z-z_0)^2 \right]  \nonumber \\
& + & U_{\rm CP}(0,z),  \nonumber
\end{eqnarray}
 with $\omega_{\bot}=\sqrt{\omega_{0 \bot}^2+\omega_{\rm CP}^2/2}$. In Eq.(\ref{eq:rotating frame equation}), $g$ is a constant related to the $s$-wave atomic scattering length $a_s$ by $g=4 \pi \hbar^2 a_s/m$. The anisotropy $\epsilon$ is expressed as
 \begin{equation}
 \epsilon= \omega_{\rm CP}^2/(2 \omega_{0 \bot}^2+\omega_{\rm CP}^2) \,.
 \end{equation}
   The normal Casimir-Polder potential $U_{\rm CP}(0,z)$ is not relevant in the following discussion. It
 slightly shifts the  sample
 vertical position, which is taken into account by a redefinition of $z_0.$
 After a stage of adiabatic approach of the plate, the rotating frame potential $U(\mathbf{r},t)$ becomes time-independent.

In order to optimize the sensitivity of our setup, we maximize the anisotropy $\epsilon$ for a given plate corrugation and a given distance $z_0$ between the sample center and the plate. Accordingly, we choose the minimum radial trap frequency providing a confinement satisfying Eqs.(\ref{eq:conditions sur les rayons}). Using the relations $\mu= \frac {1} {2} m \omega_{0 \bot,z}^{2} R_{\bot,z}$ and $\mu= \frac 1 2 \hbar \overline{\omega} \left[ 15 N a /(\hbar/ m \overline{\omega})^{1/2} \right]^{2/5}$ with $\overline{\omega}=\omega_{0\bot}^{2/3} \omega_{0z}^{1/3}$ between the Thomas-Fermi radii $R_{\bot,z}$, chemical potential $\mu$ and trap frequencies $\omega_{\bot,z}$, one finds readily
\begin{equation}
\label{eq:minimum radial frequency}
\omega_{0 \bot}(d,z_0,N)= \frac {16 \sqrt{15} \hbar (N a_s)^{1/2}} {m c^{1/2}_{\!/\!/} c^2_{\bot}} \frac {1} {\sqrt{z_0} d^2} \,.
\end{equation}
This equation suggests to use a relatively dilute condensate, for which a weak radial trapping can balance the repulsive interactions and achieve the desired confinement. The pancake-shaped samples considered is this Letter are thus not rigorously in the Thomas-Fermi regime, i.e. their transverse radius $R_{\bot}$ is comparable to the the harmonic trap scale $R_{\omega}=\sqrt{\hbar/(m \omega)}$. However, the transverse kinetic energy is much smaller than the interaction energy, so that the required trapping frequency is given by Eq.(\ref{eq:minimum radial frequency}) with a good approximation.

Let us now calculate the anisotropies that can be attained through the Casimir interaction between the condensate and the grating.
 For the low angular frequencies used here, of the order of the transverse trapping frequency and thus of a few dozens of hertz, dynamical Casimir and non-contact quantum friction effects are negligible~\cite{DiegoPauloQuantumFrictionReview}. 
 For separation distances in the micrometer range, the time it takes for
light to travel between the grating and the atom is of the order of $10^{-14}\,{\rm s}$
 and hence much shorter than the time scale associated to the grating rotation. 
 Thus, the atom interacts with the instantaneous angular position of the plate (on the other hand retardation is very important as far as charge fluctuations 
 in both atom and grating are concerned). 
 Therefore in the rotating frame the potential
  is given by the static potential, which we  compute by following the scattering approach~\cite{Lambrecht06,AnaNonPerturbativeCasimir}. The atom-surface potential is written in terms of the
 dynamic atomic polarizability $\alpha$ evaluated at imaginary frequencies $i\xi$ and of the reflection operator
 ${\cal R}_S$
 describing non-specular diffraction by the grating \cite{Messina09}.

To calculate  ${\cal R}_S$
for a periodic grating, we employ the Rayleigh basis for the fields propagating inside the homogenous  regions corresponding to the bulk material
 medium ($z>0$) and to
  the empty space below the grating ($z<-a$). The fields in the  modulated region of thickness $a$ are obtained by solving coupled differential equations. The reflection matrix elements $\langle j, \sigma | {\cal R}_S(k_x,k_y,\xi)| j', \sigma'\rangle $ (with $k_x$ varying in the first Brillouin zone $[-\pi/d,\pi/d]$ and
 $\sigma$ and $j$ representing polarization and
 diffraction order respectively) are then obtained by matching
 the expansions in the three regions across the boundaries at $z=0$ and $z=-a$.  The potential at position $(x,y,z)$ is written as
  \begin{eqnarray} \label{final}
 && U_{\rm CP}(x,z) = \frac{\hbar}{\epsilon_o c^2} \int_0^{\infty} \frac{d \xi}{2 \pi} \int_{-\infty}^{\infty} \frac{ d k_y}{2 \pi}  \int_{-\pi /d}^{\pi /d} \frac{ d k_x}{2 \pi} \nonumber \\
&& \times \sum_{j,j'}  \frac{\xi^2} {2 \kappa_{j'}} \alpha (i \xi)  \; e^ {2 \pi i (j-j') x/d}  \;  e^{-(\kappa_j + \kappa_{j'}) |z|}  \\
&& \times \sum_{\sigma, \sigma'} \langle j, \sigma | \mathcal R_S (k_x, k_z, \xi) | j', \sigma' \rangle \; \hat {\boldsymbol {\epsilon}}_ {\sigma}^- (\mathbf {k'}, i \xi) \cdot \hat {\boldsymbol {\epsilon}}_ {\sigma'}^+ (\mathbf {k'}, i \xi) \,, \nonumber
\end{eqnarray}
with $\kappa_j= \sqrt{\xi^2/c^2+ (k_x+2\pi j/d)^2+k_y^2}.$
 The unit vectors $ \hat {\boldsymbol {\epsilon}}_ {\sigma}^{\pm}$  provide the direction of the electric field propagating upwards/downwards with
 polarization $\sigma=\mbox{TE or TM}$.

 To calculate $\omega_{\rm CP},$ we first differentiate $U_{\rm CP}(x,z)$ twice with respect to $x$ and then evaluate the resulting expression at $x=0$ (middle of the plateau)  numerically. We consider a silicon rectangular (lamellar) grating with amplitude $a=4\,\mu{\rm m},$ period $d= 30\,\mu{\rm m}$
 and gap width $d/2$ (see Fig.~1). The corresponding polarizability function $\alpha(i\xi)$ is provided by Ref.\cite{Derevianko}, whereas the electric permittivity for intrinsic silicon
  at the imaginary frequency axis, required
   to compute the reflection matrix ${\cal R}_S$ in (\ref{final}), is calculated from
data at real frequencies obtained from Ref.~\cite{Palik} with the help of  a suitable  Kramers-Kronig relation~\cite{Lambrecht00}. We choose a sample of $N=100$ $\:^{87}\mbox{Rb}$ atoms, which is both sufficient to detect vortices with state-of-the art techniques and compatible with a weak radial confinement.

\begin{figure}[t]
\centering
\includegraphics[width=8cm]{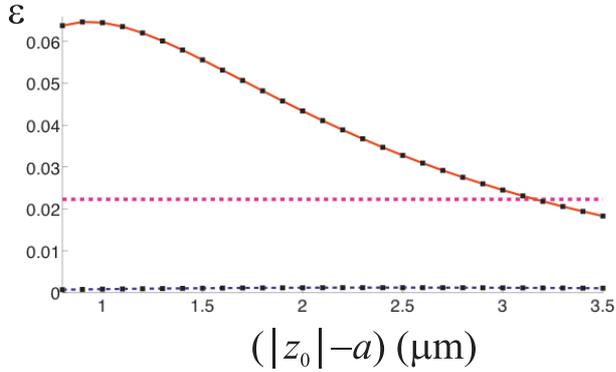}
\caption{(Color online) Anisotropy induced by the exact (full red line) and PWS (blue dashed line) Casimir-Polder potential
as a function of the distance between the plate and the sample. We consider a silicon grating with period  $d=30 \,\mu{\rm m}$ and amplitude $a=4 \,\mu{\rm m}$ and
a pancake-like condensate of $N=100$ Rb atoms harmonically confined with a transverse radius $R_{\bot}=d/4$ and a longitudinal radius of $R_z=0.2 |z_0|$. The horizontal dashed line represents the minimal anisotropy for which vortices have been produced experimentally.}
\label{fig:casimir anisotropy}
\end{figure}

 In Fig.~\ref{fig:casimir anisotropy}, we plot the anisotropy $\epsilon$ as a function of the separation distance, with $\omega_{0 \bot}$ computed from
 (\ref{eq:minimum radial frequency}) with $c_{\bot}=1$ and $c_{\!/\!/}=0.2.$ We sketch the values obtained with the non-perturbative scattering method and with the standard PWS approach. These results can be compared with the regime of vortex nucleation in condensates with quadratic rotating potentials, which have been extensively studied theoretically~\cite{RMPFetter} and experimentally~\cite{Madison01,Hodby02}. These studies determine a well-defined region in the parameter space
 of   anisotropy $\epsilon$ and angular frequency $\Omega$ leading to the nucleation of vortices when a specific route is followed.
Here we propose to follow the procedure of adiabatic increase of the anisotropy~\cite{Madison01} by slowly decreasing the distance between the rotating plate and the trapped sample until vortices emerge.
  Since the angular frequency of the plate can be tuned at will, we focus on the required anisotropy $\epsilon$ which limits the separation distance at which vortices can be nucleated. Quantized vortices have been obtained experimentally with anisotropies  as low as $\epsilon_{\mbox{min}}=2.23\%$~\cite{Hodby02}. Even if these studies have been conducted in the Thomas-Fermi regime and are thus not quantitatively applicable to the dilute samples considered here, they still provide qualitative bounds which strongly support the emergence of vortices in the present setup. Fig.~\ref{fig:casimir anisotropy} shows indeed that the anisotropy induced by the Casimir-Polder interaction, as computed with the scattering method, is above the experimental minimum $\epsilon_{\mbox{min}}$ (dashed horizontal line) for distances up to $3 \,\mu{\rm m}$. The pairwise approach predicts an anisotropy $\epsilon_{\mbox{pws}}$  well below the exact theoretical value $\epsilon_{\mbox{scat}}$ and below the experimental threshold for vortex nucleation $\epsilon_{\mbox{min}}$. At a distance $|z_0|=3 \: \mu \mbox{m}$, one finds indeed $\epsilon_{\mbox{pws}} / \epsilon_{\mbox{scat}} \sim \epsilon_{\mbox{pws}} / \epsilon_{\mbox{min}} \simeq 5 \: \% $. Thus, the emergence of vortices cannot be explained by the naive picture of a rotating additive atom-surface potential. 

To conclude, we have presented evidence that the electromagnetic
 quantum vacuum fluctuations in the neighborhood of a rotating grating can transfer angular momentum and induce quantum vortices in a condensate placed a few microns from the plate.
 In the setup we have analyzed, PFA would predict a flat Casimir-Polder potential in the region of the condensate, whereas the PWS treatment predicts a distortion which is far too small to nucleate vortices in the sample. The appearance of vortices thus results from a non-trivial geometry effect on vacuum fluctuations, which is brought into evidence thanks to the superfluid nature of the sample. For the experimental parameters considered in our numerical example, significant anisotropies - higher than those required to obtain vortex nucleation in the Thomas-Fermi regime - can be achieved for distances up to $3\,\mu{\rm m}$.  One can also explore the quantum state engineering of other BEC modes by tayloring
  quantum vacuum fluctuations with different non-trivial plate geometries. Last, we underline that we have proposed the quantum analogue of a rotating bucket which does not touch the sample: this setup illustrates, through the nucleation of quantized vortices, how quantum fluctuations may
  mimic contact interactions at a distance and provide contactless transfer of angular momentum.
  
The authors would like to thank Jean Dalibard and Peter Rosenbuch for enlightening discussions.
This work was partially supported by CAPES-COFECUB, CNPq, DARPA, DGA,
ESF Research Networking Programme CASIMIR and  FAPERJ-CNE.

\quad

\end{document}